# Interaction of optical pulse – tumor using finite element analysis

Xianlin Song [a, #, *], Ao Teng [b, #], Jianshuang Wei [c, d, #], Hao Chen [e], Yang Zhao [a], Jianheng Chen [a], Fangwei Liu [a], Qianxiang Wan [a], Guoning Huang [a], Lingfang Song [f], Aojie Zhao [a], Bo Li [a], Zihao Li [a], Qiming He [a], Jinhong Zhang [a]

[a] School of Information Engineering, Nanchang University, Nanchang, China;
[b] Institute for Advanced Study, Nanchang University, Nanchang, China;
[c] Britton Chance Center for Biomedical Photonics, Wuhan National Laboratory for Optoelectronics-Huazhong University of Science and Technology, Wuhan, China;
[d] Moe Key Laboratory of Biomedical Photonics of Ministry of Education, Department of Biomedical Engineering, Huazhong University of Science and Technology, Wuhan, China;
[e] School of Software, Nanchang University, Nanchang, China;
[f] Nanchang Normal University, Nanchang, China;
Email: *songxianlin@ncu.edu.cn

## Abstract

Photoacoustic imaging is an emerging technology based on the photoacoustic effect that has developed rapidly in recent years. It combines the high contrast of optical imaging and the high penetration and high resolution of acoustic imaging. As a non-destructive biological tissue imaging technology, photoacoustic imaging has important application value in the field of biomedicine. With its high efficiency bioimaging capabilities and excellent biosafety performance, it has been favored by researchers. The visualization of photoacoustic imaging has great research significance in the early diagnosis of some diseases, especially tumors. In photoacoustic imaging, light transmission and thermal effects are important processes. This article is based on COMSOL software and uses finite element analysis to construct a physical model for simulation. Through laser pulses into the stomach tissue containing tumor, the physical process of light transmission and biological heat transfer was studied, and a photothermal model composed of two physical fields was built, and finally a series of visualization graphics were obtained. This work has certain theoretical guiding significance for further promoting the application of photoacoustic imaging in the field of biomedicine.

## 1. Introduction

Photoacoustic Imaging (PAI) is an emerging non-invasive and non-ionizing biomedical imaging technology based on photoacoustic effect developed in recent years[1] [2] [3] [4] [5]. When the laser pulse reaches the biological tissue, the biological tissue absorbs the light energy, generates an instantaneous temperature rise, adiabatic expansion, and finally generates an ultrasonic signal. This ultrasound signal carries the relevant information of the tissue, which is received and analyzed to reconstruct the structure and function of the biological tissue[6] [7] [8]. Photoacoustic imaging combines the high contrast of pure optical imaging and the high resolution of pure ultrasound imaging, and can obtain high-resolution and high-contrast structural information. In addition, photoacoustic imaging is a non-destructive biological tissue imaging technology.

Gastric carcinoma is the most common malignant tumor of the stomach. Gastric carcinoma originates from the epithelium of the gastric mucosa and ranks first in the incidence of various malignant tumors in China. The vast majority of gastric cancers are adenocarcinomas, with no obvious symptoms in the early stage. They are often similar to the symptoms of chronic gastric diseases such as gastritis and gastric ulcers, and are easily overlooked. Therefore, the current early diagnosis rate of gastric cancer is still low. At present, the medical diagnosis methods for gastric cancer mainly include X-ray, gastroscopy, abdominal ultrasound, CT and positron emission imaging, and tumor markers. These detection methods are mainly for gastric cancer with obvious clinical symptoms, and the ability to diagnose early, especially for small lesions, is weak. And early diagnosis plays a vital role in patient treatment. Photoacoustic imaging has the advantages of high sensitivity, high resolution, deep imaging, and non-invasiveness. Research on the visualization of photoacoustic imaging can provide reliable theory and data support for early detection and diagnosis of gastric tumors.

The overall process of photoacoustic imaging can be summarized as the tissue absorbs light energy-produces thermal expansion-emits ultrasonic signals, including four physical processes of light transmission, biological heat transfer, solid mechanics and pressure acoustics. The four physical processes can all be described by partial differential equations. If an analytical solution of the equation is required, the idealization assumption is very demanding. For general engineering and mechanics problems, it is difficult to satisfy idealized assumptions, the solution process is complicated and cumbersome, and it is difficult to obtain analytical solutions. The problem can only be solved by numerical solutions. Finite element analysis (FEA, finite element analysis) refers to the use of mathematical approximations to simulate real physical systems, which can be used to solve partial differential equations.

In this chapter, a two-dimensional rectangular model is used to simulate a partial cross-section of the gastric tumor. The diffusion equation or the Helmholtz equation describes the transmission of laser in the brain tissue. The biological heat transfer

# equally contributed to this work

equation describes the temperature distribution of gray matter and blood vessel.

## 2. Model of gastric tumor using FEA

### 2.1. Geometry

As the digestive organ of the human body, the stomach is actually an organ similar to an elastic container, just like a balloon. As food enters the stomach, the stomach will expand accordingly, and the thickness of the stomach wall is 2-7 mm. In constructing the gastric tumor model, we used a 14 mm × 9 mm rectangle to represent the cross section of the local gastric tissue. A circle with a radius of 0.8 mm was set as the tumor in the cross section of the gastric tissue. The water layer was represented as 14 mm × 3 mm rectangle. A point-shaped laser pulse is used as the light source. The schematic diagram of the geometric model is shown in Figure 1(a).

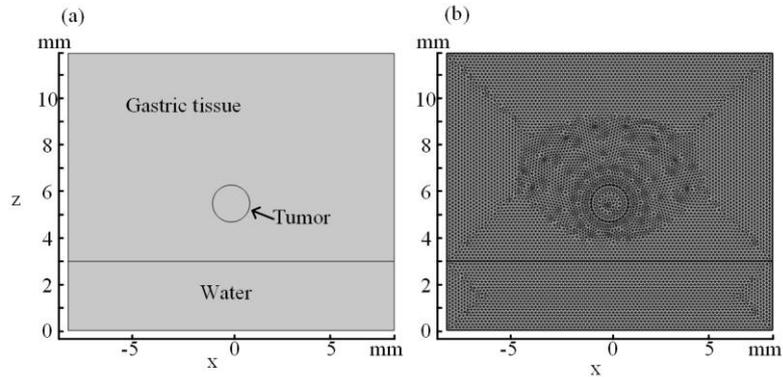

Figure 1. (a) Model geometry. (b) The division of finite elements

The key to the use of finite element analysis methods is the division of finite elements. The finer the division, the more accurate the results obtained, but the amount of calculation also increases geometrically. Therefore, in the process of simulation, it is necessary to weigh the number of finite element divisions. In addition, the rationality of the grid, that is, the quality of the grid, is directly related to the accuracy of the calculation. If the quality of the grid is too poor, the calculation may even be impossible. Therefore, when dividing the mesh, the final mesh quality should meet certain indicators. Especially in key structures, strict and reasonable division should be ensured.

In this simulation, the maximum element size of the finite element is 0.2 mm, the minimum element size is $8.6 \times 10^{-2}$ mm, the maximum element growth rate is 1, the curvature factor is 0.2, and the resolution of the narrow area is 1. The geometric model is divided, as shown in Figure 1(b).

### 2.2. Light propagation

The study of optical transmission adopts the partial differential equation in the form of coefficients in the COMSOL software, and its expression is as equation (1):

$$e_a \frac{\partial^2 u}{\partial t^2} + d_a \frac{\partial u}{\partial t} + \nabla \cdot (-c\nabla u - \alpha u + \gamma) + \beta \cdot \nabla u + au = f \tag{1}$$

Set the coefficient c in Equation (1) as the diffusion coefficient D, the coefficient a as the light absorption coefficient $\mu_a$, and the coefficient as n/c, where n is the refractive index of the corresponding tissue, and c is the speed of light in vacuum; The coefficients $e_a, \alpha, \beta, \gamma$ are set to 0, and $f$ is used as the input source. The equation in COMSOL can simulate the laser transmission in biological tissues. A lot of research has shown that the Gaussian function can be used to approximate the laser beam emitted from the optical fiber, so the source term $f$ in equation (2) is set as:

$$f = \frac{1}{4\pi^{1.5}} \frac{W_p}{\tau_p} \exp(-\frac{4 \cdot (t - \tau_c)^2}{\tau_p^2}) \delta(r - r_0) \tag{2}$$

Where, $W_p = 4 \, mJ/cm^2$ is the energy of the Gaussian laser pulse, $\tau_p = 10 \, ns$ is the pulse width, $\delta(r - r_0)$ represents the laser source at point $r_0$, $\tau_c = 30 \, ns$ represents the moment of laser pulse emission. The optical parameters of brain tissue and water are listed in **Table 1**. This module uses the Helmholtz equation [9] given in Equation (1) to solve the fluence rate $u$.

**Table 1.** Gastric tumor model with optical properties.

|  | $\mu_a (mm^{-1})$ | $\mu_s (mm^{-1})$ |
|---|---|---|
| Gastric tissue | 0.015 | 4.75 |
| Tumor | 0.742 | 0.9 |
| Water | 0.006 | 0.001 |

### 2.3. Biological heat transfer

Under laser irradiation, a certain regular light distribution will appear in biological tissues. Biological tissue absorbs laser energy and generates heat energy on the illuminated surface and inside. Due to the absorption of light energy, the temperature in the tissue changes. This kind of biological heat transfer simulation can be carried out with the biological heat transfer module of COMSOL Multiphics, as shown in equation (3).

$$\rho C \frac{\partial T}{\partial t} = \nabla \cdot (k\nabla T) + \rho_b c_b w_b [T_{art} - T] + Q_r \tag{3}$$

Where, $T_{art}$ is the arterial blood temperature, $\rho$ is the density of the tissue, $T$ is the tissue temperature, $\rho_b$ is the blood temperature, $c_b$ is the blood specific heat, $w_b$ is the blood perfusion rate, $C$ is the specific heat capacity, and $k$ is the thermal conductivity. Set the initial temperature of the tissue to 310.15 K. Since the blood vessels of the stomach wall tissue are relatively few and very small, the influence of blood is not considered in the simulation, thus, $w_b = 0$. $Q_r$ is the heat source term caused by laser irradiation, $Q_r = \mu_a \cdot u$.

This equation is the most basic equation for the heat transfer problem in the action of laser and biological tissue. It is applicable to most models of the interaction between laser and biological tissue. In the action of laser and biological tissue, the time scale of the action is very small, so it is usually does not consider the body temperature regulation mechanism. The thermal properties of gray matter and blood vessel are listed in Table 2.

Table 2. Thermal Properties.

|  | Density, $\rho$ ($kg/m^3$) | Specific Heat Capacity, $C$ ($J/kg \cdot K$) | Thermal Conductivity, $k$ ($W/m \cdot K$) |
| --- | --- | --- | --- |
| Gastric tissue | 998 | 840 | 0.7 |
| Tumor | 1000 | 4181.3 | 1.1 |

## 3. RESULTS

### 3.1. Visualization of optical energy deposition

According to the set parameters, the laser pulse is emitted at 30 ns, and the obtained optical energy distribution is shown in Figure 2. It can be seen from Figure 2(a) and Figure 2(b) that the impact of tumors on the distribution of light energy is very obvious. If there is no tumor, assuming that all parts of the stomach tissue are uniform, the light energy attenuation should be close to the same. In other words, the light intensity on the same horizontal line on the z-axis should be approximately the same. However, when a tumor is contained, it can be clearly seen from the above-mentioned visual graph that the light energy at the tumor is significantly reduced, that is, most of the photons are absorbed by the tumor, and a small part of the photons are absorbed in the area without the tumor. In addition, it can be seen from the figure that the tumor also affects the light energy distribution around it.

Figure 2(c) shows the light intensity distribution in the lateral direction. A horizontal line is taken on the z-axis through the center of the tumor. The light energy distribution on this horizontal line at different times (26 ns, 28 ns, 30 ns, 32 ns, 34 ns) is achieved. It can be seen that the light energy at the tumor is significantly lower than other places, and the light energy at the center of the tumor reaches the lowest value. Due to the symmetry of the tumor model, the distribution of lateral light intensity is also symmetric. The groove reflects the size of the tumor in this lateral direction.

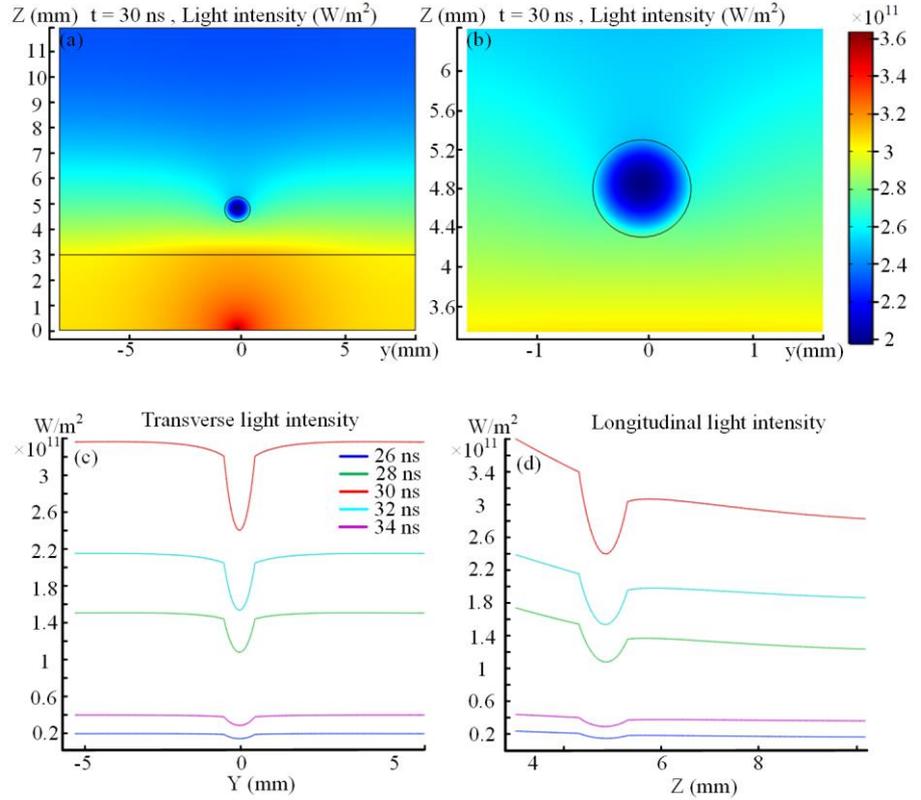

**Figure 2.** The fluence rate. (a) 30 ns. (b) Close up image of light energy at 30 ns. (c) shows the fluence rate of different moments at 26 ns, 28 ns, 30 ns, 32 ns, 34 ns of z = 4.8 mm. (d) shows the fluence rate of different moments at 26 ns, 28 ns, 30 ns, 32 ns, 34 ns of y = 0 mm.

Figure 2(d) is the light intensity distribution in the longitudinal direction. A vertical line is taken on the y-axis through the center of the tumor. The light energy distribution on this longitudinal line at different times (26 ns, 28 ns, 30 ns, 32 ns, 34 ns) is achieved. The distribution of light energy along the longitudinal line also has grooves. This indicates that a large amount of light energy is absorbed at the tumor, where the remaining light energy is less than the remaining light energy at the stomach tissue. The light energy at the tumor is significantly lower than other places, and the light energy at the center of the tumor reaches the lowest value. The light energy itself attenuates as the light propagates in the tissue, resulting in asymmetrical light intensity distribution. The light intensity on the tumor surface near the light source is higher than that on the far side.

### 3.2. Visualization of instantaneous temperature rise

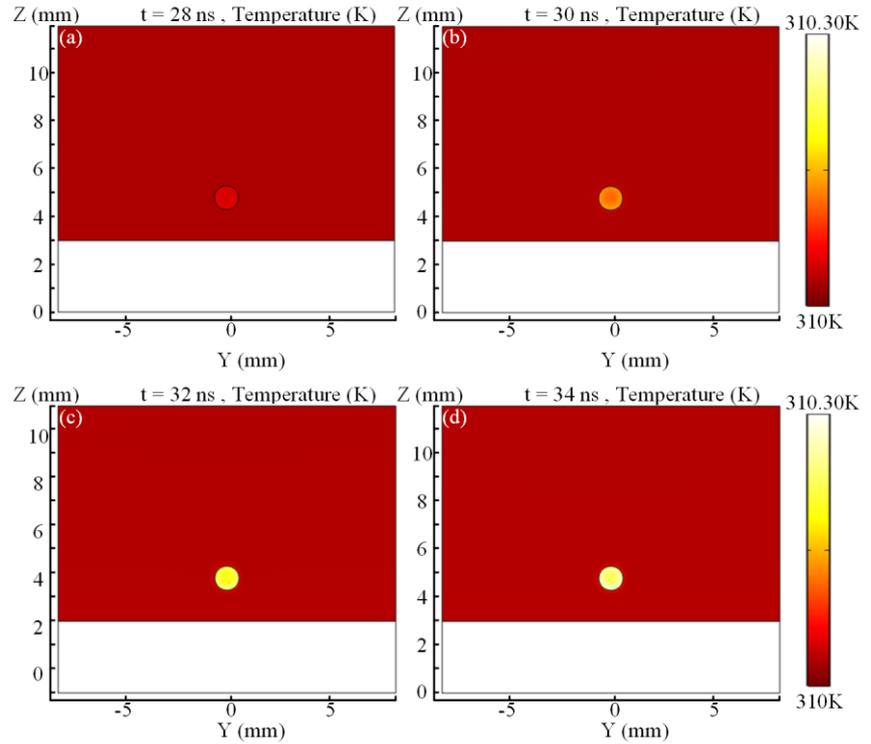

**Figure 3.** Temperature distribution at different moment. (a) 28 ns. (b) 30 ns. (c) 32 ns. (d) 34 ns.

It can be seen from Figure 3 that the influence of the tumor on the temperature distribution is also very obvious. It is obvious from the figure that the temperature of the tumor area is significantly higher than that of the stomach tissue. This is because most of the photons are absorbed by the tumor, and light energy is converted into heat energy in the tumor. The peak energy of the laser pulse is 30 ns. Since it takes a certain time to convert from light energy to thermal energy, the temperature change is slightly lagging behind the change of light energy.

In order to be able to see the temperature distribution of the tumor and the gastric tissue more clearly, a horizontal line and a vertical line running through the center of the tumor were selected in the model to study the temperature distribution in these two directions, as shown in Figure 4. Figure 4(a) and Figure 4(b) is a close up image of temperature of tumor at 28 ns and 32 ns, respectively. In Figure 4(c), the temperature at the tumor is significantly higher than that of the surrounding gastric tissue. This is because most of the light energy is absorbed by the tumor and converted into heat energy, and the temperature at the center of the tumor is significantly lower than the temperature on the surface of the tumor. The surface temperature of the tumor gradually increased, forming a concave pattern. Because the tumor is symmetrical in the lateral direction, the temperature distribution is also symmetrical. In Figure 4(d), the longitudinal temperature distribution shows that the temperature at the tumor is significantly higher than the surrounding gastric tissue. Since the light energy itself attenuates as the propagation distance increases, the light energy absorption on the side far from the light source is less than that on the side close to the light source. Therefore, the tumor surface temperature on the side closer to the light source is higher than that on the far side.

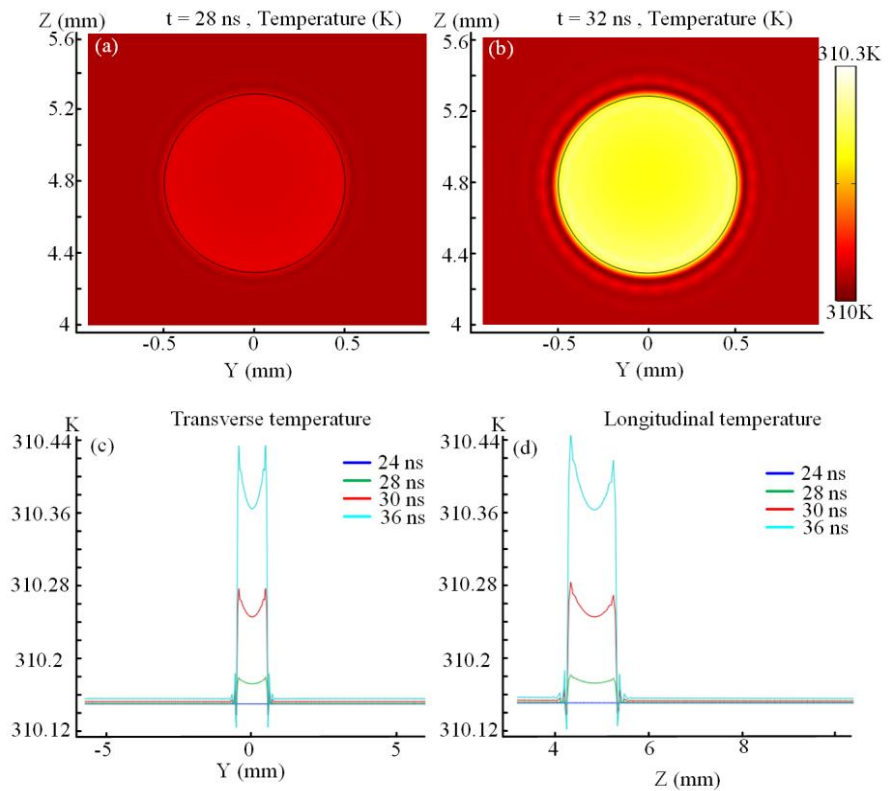

**Figure 4.** (a) Close up image of temperature at 28 ns. (b) Close up image of temperature at 32 ns. (c) The lateral temperature distribution. (d) The longitudinal temperature distribution.

## 4. Conclusion

In summary, we proposed visualization of interaction of laser pulse – gastric tumor by COMSOL based on finite element analysis. Through laser pulses into the stomach tissue containing tumor, the physical process of light transmission and biological heat transfer was studied, and a photothermal model composed of two physical fields was built, and finally a series of visualization graphics were obtained. This work has certain theoretical guiding significance for further promoting the application of photoacoustic imaging in the field of biomedicine.

## Conflicts of Interest

The authors declare no conflicts of interest regarding the publication of this paper.